\documentclass[a4paper,11pt]{article}
\usepackage{pos}
\usepackage[export]{adjustbox}

\usepackage[caption=false]{subfig}
\usepackage{graphicx}

\renewcommand{\logo}{\relax}

\title{Diffraction and elastic scattering at the LHC}
\author*[a,b]{Anna Fehérkuti}
\onbehalf{for the ATLAS, CMS and LHCb Collaborations}

\affiliation[a]{Department of Atomic Physics, Eötvös Loránd University,\\
    Pázmány Péter stny. 1a, Budapest 1117, Hungary}
\affiliation[b]{Wigner Research Centre for Physics,\\
    Konkoly-Thege Miklós út 29-33, Budapest 1121, Hungary}
\emailAdd{anna.feherkuti@cern.ch}

\abstract{Several LHC experiments exploit the high acceptance in the forward region. Most of them use the possibility of measuring the intact proton in an elastic event. This approach can enhance the purity of selecting jet-gap-jet events, leading to refined limits on photon-induced processes or to a better study of pion pair production. Diffraction and elastic scattering plays also the main role when measuring the $\varrho$-parameter or the nuclear slope in total cross-section measurements.

In this talk a summary of the latest results on this topic from the ATLAS, CMS and LHCb experiments has been presented.}

\FullConference{The Eleventh Annual Conference on Large Hadron Collider Physics (LHCP2023)\\
 22-26 May 2023\\
 Belgrade, Serbia\\}

\begin{document}
\maketitle
\section{Introduction}
Diffraction and elastic scattering covers a vast range of different studies. It contains total cross-section measurements (in which our knowledge about the so-called $\varrho$-parameter and the nuclear slope can be also tuned), examination of jet-gap-jet events or pion pair production. One can also examine photon-induced processes, coherent charmonium production, central exclusive production (CEP) of $t\Bar{t}$, searching for missing energy, light-by-light (LbL) scatterings at high mass, axion-like particles (ALPs), or setting new limits on vector boson quartic anomalous couplings.

The experiments in scope are ATLAS, CMS, TOTEM and LHCb. The forward subdetectors used for proton-tagging are the ALFA and ATLAS Forward Detectors (AFP), the Precision Proton Spectrometer (PPS), the Roman Pots (RPs) and the high-rapidity shower counters for LHCb (HeRSCheL), respectively.

Most of the results presented use proton-proton data at $\sqrt{s} = 13$~TeV, except where it is indicated not to be the case.

\section{Results on diffraction and elastic scattering at the LHC}
\begin{figure*}
    \centering
    \begin{minipage}[t]{0.5\textwidth}
        \subfloat[]{\includegraphics[width= 0.8\linewidth,valign=t]{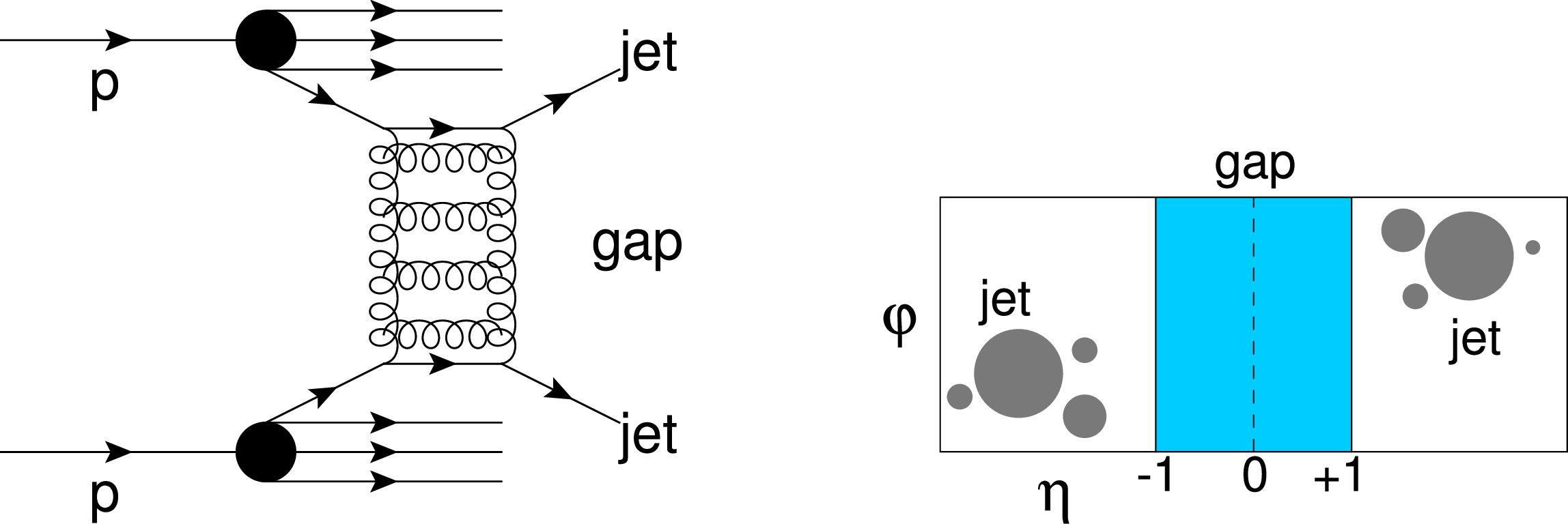}}\\
        \subfloat[]{\includegraphics[width= 0.8\linewidth]{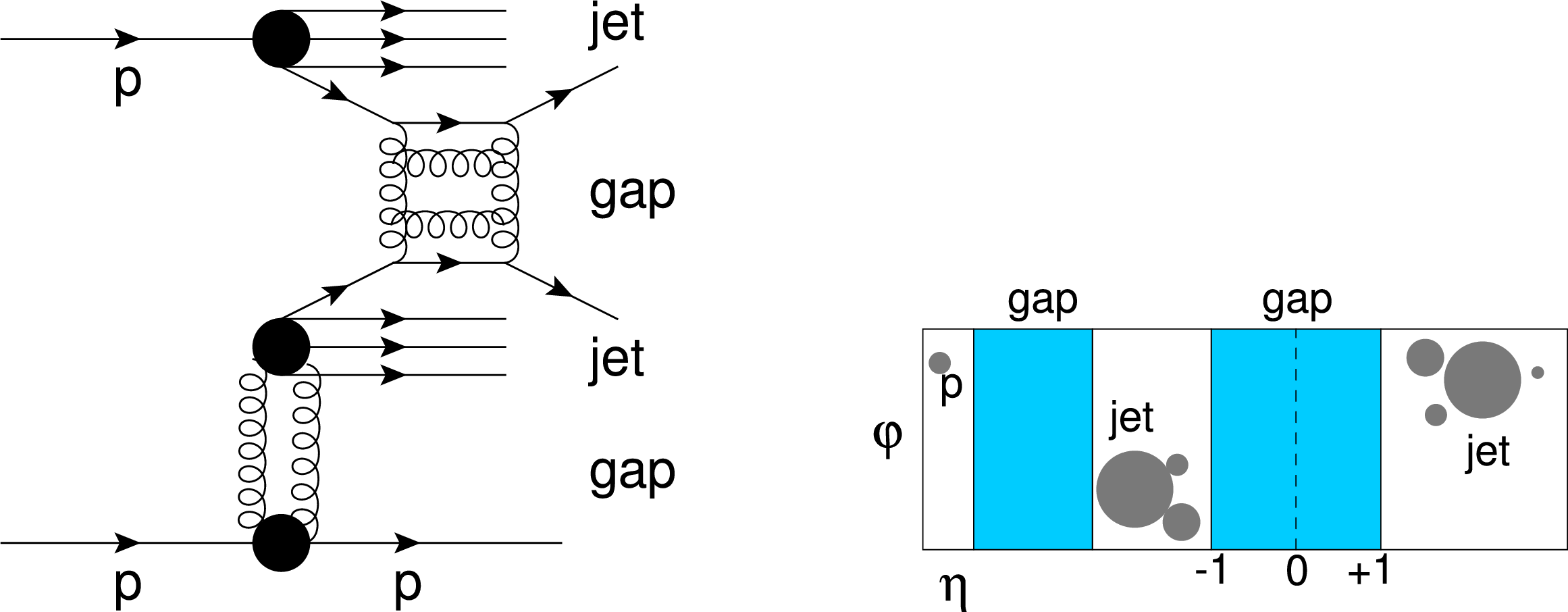}}\\
    \end{minipage}%
    \begin{minipage}[t]{0.5\textwidth}
        \subfloat[]{\includegraphics[width= 0.83\linewidth,valign=t]{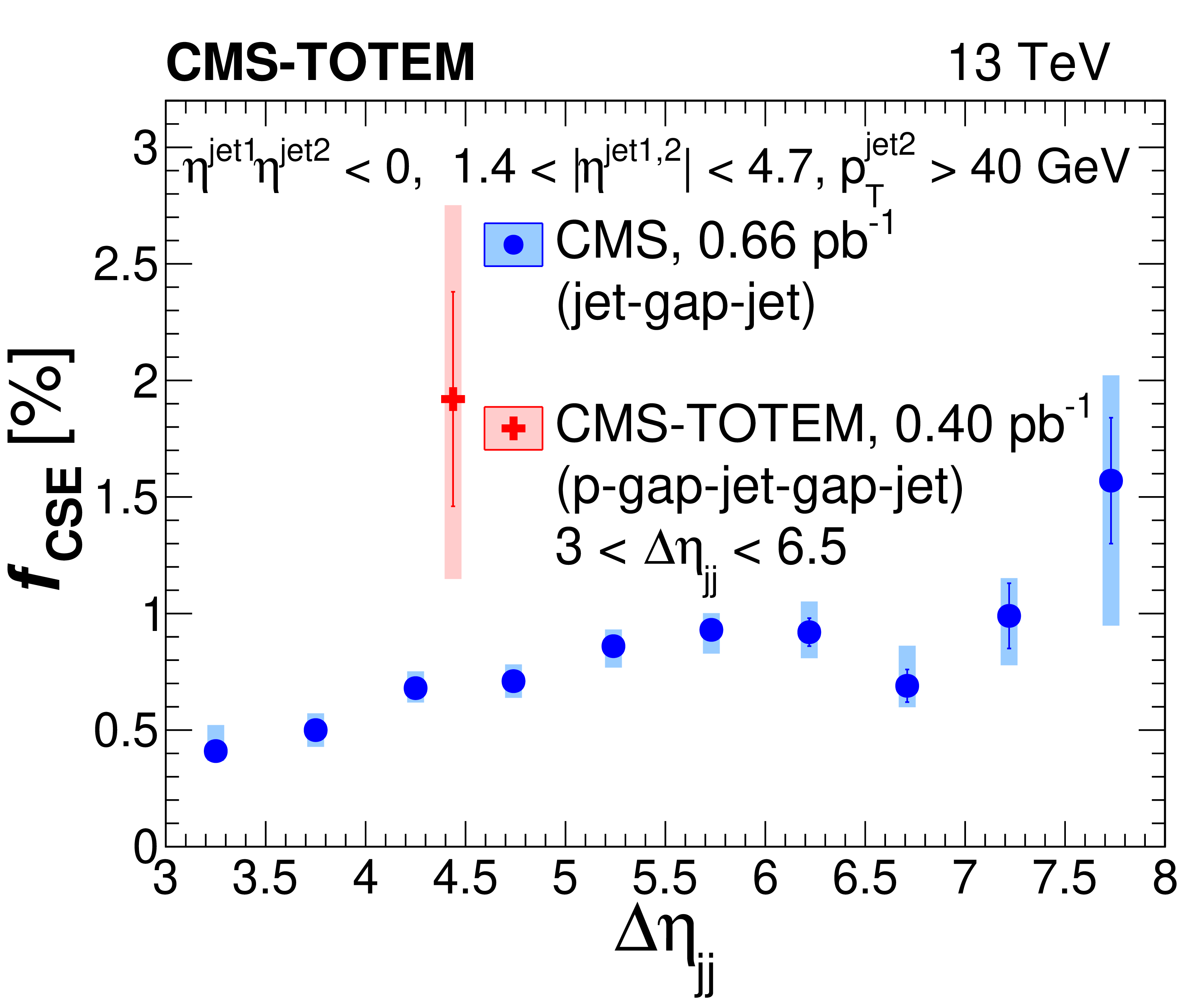}}\\
    \end{minipage}  
    \caption{Selection for (proton-gap-)jet-gap-jet events (pgjgj and jgj cases,~\cite{SMP-19-006}). Diagrams and geometry of the corresponding observed particles in the $\eta-\phi$ plane are shown on the left for both cases (a, b). Results obtained are plotted on the right (c) in blue for jgj and in red for pgjgj at the mean $\Delta\eta_{jj}$ point.}
    \label{fig:pGJGJ}
\end{figure*}

Hard color-singlet exchange in the Balitsky-Fadin-Kuraev-Lipatov (BFKL) framework is described by pomeron-exchange. The validity of the method, using resummation based on the large logarithms multiplying the strong coupling constant in the perturbation expansion, can be best studied on the so-called jet-gap-jet events, where a bigger pseudorapidity region remains unpopulated between the two jets (gap) due to the geometry of the quark-level process (see figure~\ref{fig:pGJGJ}). Defining the fraction of jet-gap-jet events as $f_{\mathrm{CSE}} = \frac{N^{\mathrm{\#tracks<3}} - N^{\mathrm{\#tracks<3}}_{\mathrm{ULE}}}{N^{\mathrm{tot}}_{\mathrm{dijet}}}$, where the $<3$ restriction has to be made (as fragmentation can also produce particles arriving in the gap region) and $ULE$ stands for the underlying events, one can plot the measured values with respect to three meaningful variables and compare the trends with predictions of theoretical models. In In a CMS and TOTEM analysis~\cite{SMP-19-006}, it has been shown that the measured $f_{\mathrm{CSE}}$ follows the best (with respect to the pseudorapidity difference between the two jets, which is the most important variable) the theoretical model not only using the resummation and multi-particle interactions, but also soft color interactions. With respect to the transverse momentum of the second (not leading) jet, $f_{\mathrm{CSE}}$ seems to be constant, in accordance with previous measurements~\cite{Previous1, Previous2, Previous3, Previous4, Previous5}. Regarding the behaviour with respect to the azimuthal difference between the jets, a peak is visible around $\pi$, which can be interpreted in a way that the jet-gap-jet jets are more correlated in the transverse plane than inclusive dijets. These results already show that jet-gap-jet events provide a powerful test of the BFKL resummation, but if one considers a subsample, requesting in addition at least one intact proton on either side of CMS (resulting in the so-called proton-gap-jet-gap-jet events, also shown in figure~\ref{fig:pGJGJ}), one can further improve the selection power. For the first time CMS observed 11 such events (\cite{SMP-19-006}) obtaining very clean events, since multi-parton interactions can be suppressed this way. It also indicates that this might be the more ideal way of probing to probe the BFKL framework in the future, since the $f_{\mathrm{CSE}}$ extracted is about three times larger than that of the inclusive case (see figure~\ref{fig:pGJGJ}).

Studying photon-induced processes, such as the exclusive production of $W$/$Z$ pairs in CMS~\cite{SMP-21-014} one can define a signal region on the basis of the correlation between the $WW$/$ZZ$ system (regarding their invariant mass and rapidity) and the tagged proton information. The limits obtained on the standard model cross sections in the fractional momentum loss region corresponding to the detector acceptance can be transferred to new limits on quartic anomalous couplings. These are better constraints with respect to analyses without proton tagging for the $W$ case, while first obtained values in the $Z$ case from the exclusive channel. Searching for exclusive diphoton production at high mass using proton tagging is also of interest at CMS~\cite{CMS-PAS-EXO-21-007}. A recent study (\cite{EXO-18-014}), using full Run 2 data obtained a more stringent limit on the standard model light-by-light production cross section, as well as on the quartic photon anomalous couplings. The latter also provides new limits on the coupling strength of ALPs at high mass. ALPs are also studied at ATLAS (\cite{EXOT-2019-28}), using AFP for proton-tagging. Besides obtaining an upper limit on the ALP coupling constant, an excess with 2.5$\sigma$ local significance is observed at one mass hypothesis, but globally this is not found to be a significant deviation. The benefits of proton tagging are demonstrated in~\cite{EXO-19-009}. Since it allows to reconstruct the total mass, measuring a $Z$ or a $\gamma$, studying the missing mass distribution it provides the means to make model-independent searches for a $pp \rightarrow Z/\gamma + X$ process, where $X$ does not have to be reconstructed, and could be any particle. In~\cite{EXO-19-009} upper limits on the cross section in both the $Z$ and $\gamma$ cases are shown. Central exclusive production of $t\Bar{t}$ includes, besides photon-induced production, also pomeron-exchange. Measuring the top pair one can look at two channels as well. Both the dilepton and the lepton+jets decay modes were measured at CMS~\cite{CMS-PAS-TOP-21-007}, from which the results were then combined obtaining an upper bound on the production cross-section. To enhance the signal sensitivity a multivariate analysis (MVA) was performed, using a boosted decision tree (BDT) algorithm.

Coherent charmonium production in ultraperipheral collisions (UPCs) was examined by LHCb~\cite{LHCb-PAPER-2022-012} in lead-lead collisions at $\sqrt{s_{NN}} = 5.02$~TeV. Coherent production means that not only one nucleon (which would be the incoherent case), but the whole nucleus is taking part in the pomeron-exchange. In an incoherent case the nucleus containing the interacting nucleon falls apart, from which the remnants can be measured with the forward detectors of LHCb. Since in the coherent case the nucleus stays intact, an extra veto on forward detectors can be used for selecting the coherent events. The cross sections measured for coherent production of $J/\psi$ and $\Psi(2S)$ can be compared to theoretical predictions. Ref.~\cite{LHCb-PAPER-2022-012} shows that these are in reasonable agreement.

A search for exclusive pion pair production in correlation with detected protons was performed by ATLAS using ALFA ~\cite{STDM-2017-07}. This was the first usage of proton tagging to measure an exclusive hadronic final state. The cross section was determined in two kinematic regions - corresponding to the two ways of pairing the RPs (in the analysis) on the opposite side of the interaction point, known as the elastic and the anti-elastic arrangement. The existing physical models in scope reproduced the measured data, which can be also the result of the limited statistical precision. However, one can conclude that on the basis of the study it is not possible to either exclude or tune any of these models. On the other hand it can be said that they provide preliminary theoretical estimates.

The elastic scattering amplitude can be factorized to a nuclear- and a Coulomb Nuclear Interaction (CNI) part. The purely strong-interacting amplitude can be parametrized as $f_N(t) = (\varrho + \mathrm{i}) \frac{\sigma_{\mathrm{tot}}}{\hbar c} \mathrm{exp}^\frac{-B|t| -C|t|^2 -D|t|^3}{2}$, where the exponential part is called the nuclear slope and the $\varrho$-parameter is the fraction of the real and the imaginary part of the elastic scattering amplitude in forward directions. In optical theorem the hadronic component of the total cross section $\sigma_{\mathrm{tot}}$ is connected to the imaginary part of the scattering amplitude in the forward direction, which means a close correlation between the $\varrho$-parameter and $\sigma_{\mathrm{tot}}$. A detailed measurement on the total cross section was carried out at ATLAS~\cite{STDM-2018-08} aiming to obtain more precise values for the above-mentioned quantities. Regarding the nuclear slope different parametrizations were used (concerning only the $B$-term or more) and a data-driven fit was applied on the measured elastic cross section leading to the final result presented in~\cite{STDM-2018-08}. This value, however, is lower than the previous TOTEM result (\cite{TOTEMComparison}) causing a yet unresolved 2.2$\sigma$ tension, which most probably comes from the fact that TOTEM used a luminosity-independent way of measuring, while the one of ATLAS is luminosity-dependent. The obtained value for the $\varrho$-parameter agrees between the two measurements.

\section{Conclusions}
As a summary one can conclude that studying the jet-gap-jet events offer a powerful test of BFKL resummation, moreover if one proton has been tagged as well. Using the LHC as a photon collider, very clean events can be obtained due to measuring intact protons and produced particles in either ATLAS or CMS. Proton tagging draws us to higher selection efficiency even in hadronic production processes. Search for exclusive $\gamma\gamma$, $ZZ$, $WW$, $t\Bar{t}$ leads to best sensitivities to quartic anomalous couplings as well as to the productions of ALPs at high mass. Not only the cross-sections but even their ratio (determined for the first time) of the cross-sections between coherent $J/\psi$ and $\Psi(2S)$ production are found to be compatible with theoretical models. Regarding the total cross-section measurements, however, the commonly accepted models are in agreement with only one of the $\sigma_{\mathrm{tot}}$ or $\varrho$ measurements, while a simultaneous fit was found to give a good description of both quantities. Nevertheless, it is still a question if the low value of $\varrho$ can be attributed to the Odderon or other effects in strong interactions.

\section{Aknowledgements}
The author would like to thank the support of the OTKA grants (K128713, K128786, K146913, K146914).

\end{document}